\documentclass[psfig,referee]{aa}                                               
\input psfig.tex

\begin{document}                                                                
%-------------------- own definitions -----------------------                   
\def\et{et al.}                                                                 
\def\egs{erg s$^{-1}$}                                                          
\def\egsc{erg s$^{-1}$ cm$^{-2}$}                                               
\def\msu{M$_{\odot}$\ }                                                         
\def\kms{km s$^{-1}$ }                                                          
\def\kmsM{km s$^{-1}$ Mpc$^{-1}$ }                                              
% ------------------------------------------------                              
   \thesaurus{06         % A&A Section 6: Form. struct. and evolut. of stars    
              (03.11.1)}  % Cosmogony,                                          

   \title{The X-ray Morphology of the Lensing Galaxy Cluster Cl0024+17}         
                                                                                
   \author{H. B\"ohringer \inst{1}, G. Soucail \inst{2},
   Y. Mellier \inst{3,4}, 
    Y. Ikebe \inst{1}, P. Schuecker \inst{1} }

   \offprints{H. B\"ohringer}                                                   
                                                                                
   \institute{$^1$ Max-Planck-Institut f\"ur Extraterrestrische Physik,         
                   D-85748 Garching, Germany\\                                  
              $^2$ Observatoire Midi-Pyr\'en\'ees, Laboratoire d'Astrophysique,
                   UMR 5572, 14 Avenue E. Belin, F-31400, France\\
              $^3$ Institut d'Astrophysique, 98 bis Bd. Arago, F-75014 Paris, France\\
              $^4$ Observatoire de Paris, DEMIRM, 61 avenue de l'Observatoire, 
                   F-75014 Paris, France 
                   }

   \date{Received ..... ; accepted .....}                                       
                                                                                
   \maketitle                                                                   
                                                                                
   \markboth {X-ray Morphology of Cl0024+17}{}                                  
                                                                                
\begin{abstract}                                                                
We report the analysis of a very deep ROSAT HRI observation
on one of the most interesting, distant lensing clusters,
Cl0024+17. Using the X-ray surface brightness we analyse
the cluster morphology and constrain the gas and gravitational
mass of the cluster.
We confirm the small core radius of the mass halo                     
of $66 {-25\brack +38} h_{50}^{-1}$ kpc for this cluster 
inferred previously from              
a detailed strong lensing analysis by Tyson et al. (1998)
and Smail et al. (1997).
Using estimated gas temperatures we find a cluster mass                      
of about $3 - 14 \cdot 10^{14} h_{50}^{-1}$ M$_{\odot}$ for a 
fiducial radius of $3 h_{50}^{-1}$             
Mpc. This mass is lower than the mass implied by the weak lensing               
result of Bonnet et al. (1994) and inconsistent with a virial                   
analysis based on the high observed galaxy velocity dispersion.
The lower gravitational mass found in the present study implies,
however, a gas mass fraction             
of the cluster e.g. at $1 h_{50}^{-1}$ Mpc radius of 
$17(11-28)~ h_{50}^{-1.5}~ \%$
well consistent with the general observations in rich clusters.                 
This favours a lower mass value for the relaxed part of Cl0024+17
which could still be embedded in a larger structure achieving a
consistency with the weak lensing observations.


   \end{abstract}                                                               
                                                                                
%                                                                               
%________________________________________________________________               
                                                                                
\section{Introduction}                                                          
Cl0024+17 is one of the most interesting distant 
($z=0.39$, Gunn \& Oke 1975) galaxy clusters     
featuring gravitational lensing effects. In this cluster eight, partly very     
detailed images of a single background galaxy have been identified              
(Colley, Tyson, \& Turner 1996, 
Tyson, Kochanski, \& Dell'Antonio 1998) which allows very detailed              
modeling of the underlaying                                                    
mass distribution of the cluster center. 
It is also the cluster in which the first     
very large scale gravitational shear field has been detected and 
characterized   
(Bonnet, Mellier, \& Fort 1994) with a significant shear signal out to a radius of          
almost 3 $h_{50}^{-1}$ Mpc (the typical virial radius of very rich           
clusters).                                                                      
                                                                                
The cluster has been discovered by Humason \& Sandage (1957) and               
was one of the first targets to display the so-called                        
Butcher-Oemler effect (Butcher \& Oemler 1978). Dressler \& Gunn                     
(1982), Dressler, Gunn, \& Schneider (1985) and 
Schneider, Dressler, \& Gunn (1986)  have found that the 
cluster is very rich, but has a large number of blue galaxies, 
confirming now with a redshift survey the earlier found
Butcher-Oemler effect.
                                                                                
Gravitational arcs were discovered in the cluster by Koo (1988) and      
subsequently studied by a number of authors (Mellier et al. 1991, 
Kassiola, Kovner, \& Fort 1992, Kassiola et al. 1995, 
Smail et al. 1997, Wallington, Kochanek, \& Koo 1995, 
Colley et al. 1996, Tyson et al. 1998).             
The most impressive arc system is located at a radius of about 35 arcsec
($\sim 223 h_{50}^{-1}$ kpc)       
presumably very close to the critical radius of the cluster lens.     
The redshift of the lensed background galaxy, which is very difficult
to determine due to the lack of covenient emission lines for
the redshift of the source (e.g. Mellier et al. 1991), has recently been
measured by Broadhurst et al. (1999) to be $z = 1.675$.     

Various mass estimates have been conducted for the cluster. 
From the measured line-of-sight  
velocity dispersion of $\sigma_r = 1287 $ km s$^{-1}$ and an optical            
core radius of $168 h_{50}^{-1}$ kpc, Schneider, Dressler \& Gunn  
(1986) calculate a gravitational mass                                           
of $ 6.6 \cdot h_{50}^{-1} 10^{14}$ M$_{\odot}$ within a radius of              
$ 0.48 h_{50}^{-1}$ Mpc.                                                        
On a much larger scale Bonnet et al. (1994) find a lensing mass of about 
$2.4 - 4 \cdot 10^{15}  h_{50}^{-1} $M$_{\odot}$ within a radius 
of $3 h_{50}^{-1}$ Mpc. A recent analysis of the shear field of 
Cl0024+17 is also included in the work by van Waerbeke et al. (1997).        
Mass estimates for the central region of the cluster
are discussed in section 4. 
                                                                                
In the following we will be using a Hubble constant of $H_0 =                   
50$ km s$^{-1}$ Mpc$^{-1}$ and $q_0 = 0.5$ and indicate the scaling
of important parameters with $h_{50} = H_0/50$ km s$^{-1}$ Mpc$^{-1}$. 
For this cosmology                  
1 arcmin at the cluster redshift corresponds to a comoving                      
scale of $382 h_{50}^{-1}$ kpc. In Section 2 we describe 
the observations and the morphological analysis. Section 3
provides mass estimates and in Section 4 we discuss
these results in comparison with the lensing properties of 
the cluster.

%__________________________________________________________________             
                                                                                
\section{ROSAT Observations}                                                    
                                                                                
Cl0024+17 was observed with the ROSAT HRI in January 1994, July 1994,           
July 1995, and June to July 1996 with a total effective exposure time           
of 116.5 ksec.                                                                  
Fig. 1 shows the ROSAT HRI image of the cluster in the form of a contour        
plot. The image was divided by the exposure map, 
background subtracted, and corrected for vignetting effects.    
The image has been smoothed with a variable Gaussian filter, with a             
filter sigma varying from $\sigma = 1.4$ arcsec for the brightest to            
$\sigma = 6$ arcsec for the fainter regions, in order to provide a large
dynamical range for the display of structural features.
There are several point 
sources discussed in detail by Soucail et al. (1999).
                                                                                
Significant diffuse emission is detected from the cluster source out to a       
radius of about 1.5 arcmin ($0.57 h_{50}^{-1}$ Mpc). The total source count     
rate within a radius of 2 arcmin is $7 (\pm 0.5) \cdot 10^{-3}$ cts s$^{-1}$    
where the count rate of the closest point source with                           
$ \sim 0.9 \cdot 10^{-3}$ cts s$^{-1}$ has been subtracted.             
Assuming a temperature of 3.6 keV (this assumption is justified in 
Section 3) and considering 
the measured galactic hydrogen column density, $N_H = 4.4          
\cdot 10^{20}$ cm$^{-2}$ (Dickey \& Lockman 1990),  
this corresponds to a flux of $F_X = 3.6               
\cdot 10^{-13}$ erg s$^{-1}$ cm $^{-2}$ and a rest frame X-ray luminosity of    
$L_X = 2.4 (\pm 0.16)\cdot 10^{44} h_{50}^{-2}$ erg s$^{-1}$
in the ROSAT band (0.1 - 2.4 keV). These results are very         
insensitive to the assumed temperature, would we have adopted a                 
temperature of 7 keV for example the derived X-ray luminosity                   
would be $L_X = 2.3 (\pm 0.16)\cdot 10^{44}h_{50}^{-2}$ erg s$^{-1}$. These  
values are consistent with the X-ray data quoted in Smail et al. (1998).       
                                                                                
\begin{figure}                                                                  
\psfig{figure=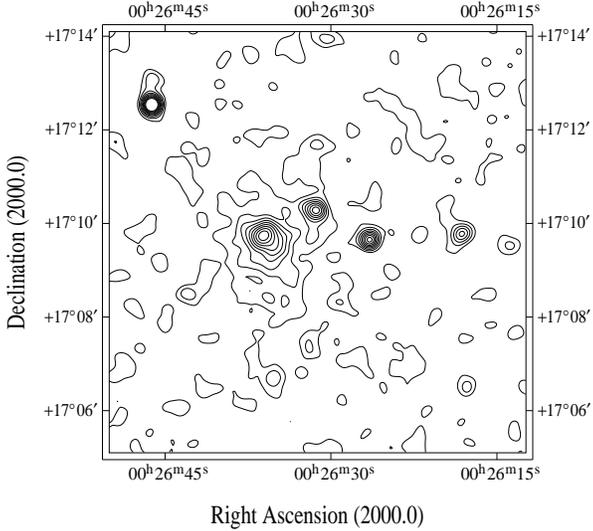,height=8cm}                                            
\caption{ROSAT HRI image of Cl0024+17. The image is background subtracted       
and vignetting corrected and has been smoothed with a variable Gaussian         
filter. See text for details.}                                                  
\end{figure}                                                                    

\begin{figure}                                                                  
\psfig{figure=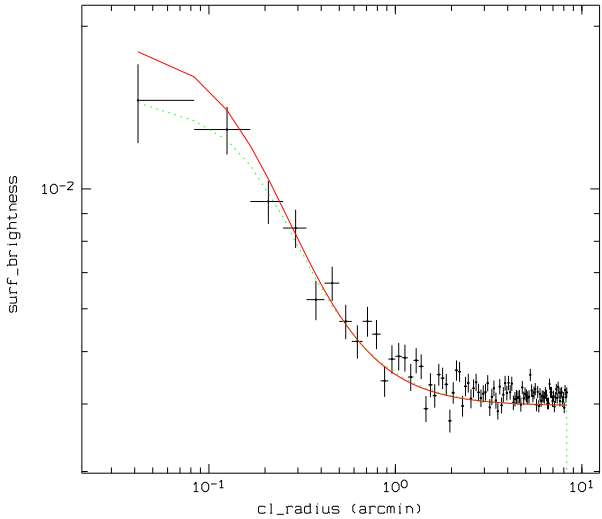,height=7cm}                                           
\caption{Surface brightness profile for the HRI image of Cl0024+17. The
photon statistical errors are given as vertical error bars.  The solid   
line shows the best fitting unconvolved $\beta $-model and the                 
dashed line shows the convolved, actually fitted profile.}                     
\end{figure}

%_________________________________One column table----------------------------  
                                                                                
   \begin{table}                                                                
   \caption{Results of the $\beta $-model fits to the surface brightness         
               profiles of the ROSAT HRI observations}                          
                                                                                
   \label{Tempx}                                                                
      \[                                                                        
         \begin{array}{llllll}                                                   
            \hline                                                              
            \noalign{\smallskip}                                                
     {\rm data}~~~~~~~~~~~~~~~~~~~~~~~~~~~~&  S_0   & {\rm core~ radius} &  
       {\rm core~ radius}~~~~~ &\beta \\
        &  {\rm cts ~s}^{-1} {\rm arcmin}^{-2} &{\rm arcsec} 
        & h_{50}^{-1}{\rm kpc} &   \\
            \noalign{\smallskip}                                                
            \hline \\                                                           
            \noalign{\smallskip}                                                
 {\rm  HRI~ uncorrected}  & 0.011 & 12.0  & 76.4  & 0.465  \\                           
 {\rm  HRI~ corrected~ for~ PSF} &  0.015 & 10.4 & 66.2 & 0.475  \\                   
            \noalign{\smallskip}                                                
            \hline                                                              
         \end{array}                                                            
     \]                                                                         
   \end{table}                                                                  
%                                                                               
%                                                                               
                                                                                
%________________________Table-END______________________________________        

We have determined an azimuthally averaged surface brightness profile           
for the HRI observation of Cl0024+17.                                           
A fit of a $\beta$-model (e.g. Cavaliere \& Fusco-Femiano 1976,
Jones \& Forman, 1984) of the form                                              
                                                                                
\begin{equation}                                                                
S(r) = S_0 \left( 1 + {r^2 \over r_c^2} \right)^{-3\beta +1/2}               
\end{equation}

\noindent                                                                       
to the data is found to provide a good description of the 
surface brightness profile. Note, however, that the fit is restricted     
to the inner $\sim 0.6 h_{50}^{-1} $ Mpc where we see significant X-ray         
emission. First we fitted the model directly to the photon data binned 
in concentric rings. Alternatively we took the smoothing effect of the HRI 
point spread function (PSF) into account by performing a 2-dimensional 
convolution of the $\beta$-models with the HRI on-axis PSF
(David et al. 1995) before fitting to the observational data. 
The cluster has a surprisingly                              
small core radius -- only $66 h_{50}^{-1} $ kpc in physical scale.              
In this case accounting for the PSF has a significant effect.               
The fitting results are summarized in Table 1 and the best fitting model        
is shown in Fig. 2 along with the observed data. In the fits the parameters
for the core radius and $\beta$ are correlated and therefore have large
individual uncertainties. For a 68\% uncertainty level we find the following
constraints for the two parameters: $\beta = (0.425 - 0.550)$ and
$r_c = (6.5\arcsec\ - 16.5\arcsec\ )$.                                

\begin{figure}                                                                  
\psfig{figure=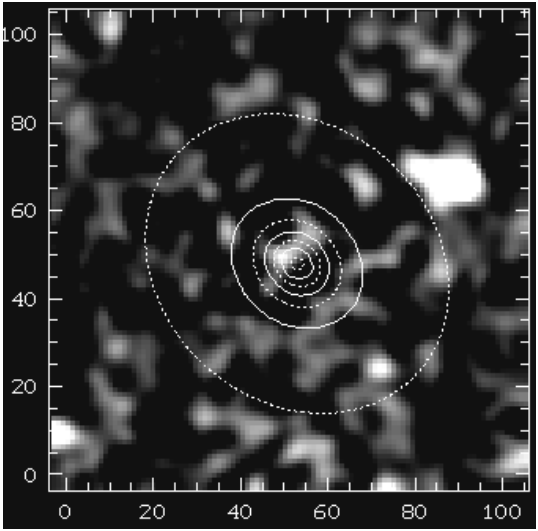,height=7cm}                                        
\caption{Residuals of the X-ray image of Cl0024+17 after subtracting           
an elliptical model from the observed cluster image.                            
The model is overplotted in the form of contour lines. The maximum               
north of the cluster core and some emission in the south are the only                 
significant residual emission regions in the cluster area.
The scale of the image is $3.33 \times 3.33$ arcmin.                               
}                                                                               
\end{figure}

As can be seen in Fig. 1, the cluster has a slight elongation in a
northeast-southwest direction. To further quantify the cluster shape we have
fitted a global elliptical model with one global slope parameter, $\beta$.
The best fitting ellipse has an orientation with a position angle of 41.5
degrees measured counter-clockwise from the north with a major axis core
radius of 15\arcsec\ and a minor axis core radius of 13\arcsec\ (yielding an ellipticity of
$\sim 15$\%). The slope parameter, $\beta = 0.48$, is well consistent with the
fit of the spherically symmetric model. Note that in this analysis the
profile was not deconvolved. The good agreement with the spherical model
allows us to base the further analysis on the spherical model,
as the effect of the ellipticity will almost
average out as shown in Neumann \& B\"ohringer (1996).

The residual image obtained by subtracting the elliptical model from the
observed cluster image (smoothed with a Gaussian of $\sigma = 4\arcsec\ $) is shown
in Fig. 3. Two significant features can be noted in this image: i) there is a
residual peak of the central maximum just north of the center of the
elliptical model and ii) there is some more faint emission in the south
than there is emission in the north (in addition to the possible faint 
point source which is located in the southern cluster area). 
Both features are easily explained as the result of a 
displacement of the central maximum with respect to the center of symmetry of
the overall cluster. It results in an imperfect subtraction of the central
maximum and the offset maximum shifts the center of the fitted ellipse
slightly north with respect to the large-scale cluster center leaving residual
emission in the southern part. The only significant trace of cluster
substructure that can be observed in the X-ray image of Cl0024+17 is this center
shift of the cluster core by about 12\arcsec\ ($\sim 75 h_{50}^{-1}$ kpc) to the
north approximately in the direction of the position angle of the ellipse
model. The two residuals, the small peak at the north of the center of
symmetry and the southern extension are about $2\sigma$ features. 
The disturbance is therefore not very large, as far as the X-ray emission can 
be traced ($\sim 0.5 h_{50}^{-1}$ Mpc). The bright residual feature
at the upper right of Fig. 3 is a point source not related to the cluster ICM
(source S1 identified in Soucail et al. 1999).

It is also important to note that the analysis of the X-ray surface brightness
profile together with an assumed gas temperature of 3.6 - 8 keV yields a
central cooling time of the gas of about 7 - 8 Gyr. This is probably larger
than the age of the cluster at the given redshift. Thus there was not
enough time to develop a steady state cooling flow and we should expect a very
marginal effect due to cooling of the gas in the cluster center.
                                                                                
\section{Cluster Mass}                                                          
                                                                                
To determine the cluster mass we need information on the cluster                
gas temperature in addition to the gas density profiles which                   
can be calculated from the observed surface brightness profiles.                 
One possibility to infer the cluster                   
temperature is to use the well known and reasonably tight                       
luminosity temperature relation for X-ray clusters as given for                  
example by Markevitch (1998) based on recent ASCA observations.                 
Using his relation (uncorrected for the luminosity effect of cooling flows),     
$(T_X/${ keV}$) = 2.34~ h_{50}~ (L_X/10^{44} h_{50}^{-1}$erg s$^{-1}$)$^{0.5}$,
we find a temperature of 3.6 keV as used above.              
                                                                                
We have constructed a range of mass profiles             
for the cluster allowing for a large temperature range from 3.6 - 8 keV
(with a best value of 6 keV)  
and a range from 3 - 4.5 keV (with a best value of 3.6 keV) as implied by the
$L_X - T_X$-relation. We are allowing               
for different shapes of the temperature profiles                   
using polytropic models with a range of $\gamma$-parameters                     
from 0.9 to 1.3, roughly accounting for the observed temperature variations.
The temperature profile for the polytropic models were normalized
such as to give the nominal emission measure weighted average temperatures. 
The results of the mass modeling are given in Table 2 and Fig. 4.
For comparison with masses estimated from a lensing analysis
two-dimensional mass profiles were also calculated assuming a cut-off radius
of $3  h_{50}^{-1}$ Mpc with results also given in Table 2. The choice
of the cut-off radius has little influence on the exact result. Taking 
for example an outer radius of  $5  h_{50}^{-1}$ Mpc, much larger than the 
expected virial radius of the cluster, the projected mass increases by
only about 25\%. 
                                                                         
   \begin{table}                                                                
   \caption{Results for the mass profile for the cluster Cl0024+17.             
The masses are given in units of $10^{14} h_{50}^{-1}$ M$_{\odot}$              
and the radii in units of $h_{50}^{-1}$ Mpc. The first set of values gives             
the result for an isothermal model with $T_x = 3.6 (3.0 - 4.5)$ keV 
and the values in brackets give the full model range for the first 
set of values. The second set of values give the corresponding
parameters for an adopted temperature of $T_x = 6 
(3.6 - 8.0)$ keV. The column labeled 2-dim. mass gives the projected
cluster mass onto the celestial sphere with an assumed outer
cut-off radius of $3  h_{50}^{-1}$ Mpc.}                                
                                                                                
   \label{Tempx}                                                                
      \[                                                                        
         \begin{array}{llllll}                                                   
            \hline                                                              
            \noalign{\smallskip}                                                
         {\rm radius}&  M_{grav}    & M_{gas}  & {\rm gas~mass~fract.} 
       & {\rm 2 dim. mass}\\    
            \noalign{\smallskip}                                                
            \hline \\                                                           
            \noalign{\smallskip}                                                
 {\rm for}~ T_x = 3.6 (3.0 - 4.5)~ {\rm keV}: & & & & \\
        & & & & \\
 0.22   &  0.4 (0.24-0.84)  & 0.036 &  9 ( 4-15)\%  & 0.57 (0.4-1.1) \\
 0.48   &  0.9 (0.7-1.5 )   & 0.15  & 17 (10-21)\%  & 1.3  (1.0-1.7) \\
 1.0    &  2   (1.5-2.5)    & 0.50  & 25 (20-33)\%  & 2.6  (1.7-3.4) \\
 3.0    &  5.7 (2.8-8   )   & 2.8   & 49 (35-100)\% & 5.7 (2.8 - 8)  \\
        & & & & \\
 {\rm for}~ T_x = 6 (3.6 - 8.0)~ {\rm keV}:  & & & &   \\
        & & & & \\
 0.22   &  0.6 (0.3-1.6)    & 0.038 &  6 (2-12)\%   &0.9 (0.48-1.8) \\
 0.48   &  1.4 (0.8-2.6)    & 0.15  & 11 (6-19) \%  & 2.2   (1.2-3) \\
 1.0    &  3.1 (1.8-4.5)    & 0.50  & 16 (11-28)\%  & 4.4   (2 - 6)\\
 3.0    &  9.4 (3.4-14.)    & 2.8   & 30 (20 - 82)\%& 9.2 (3.4 - 14.)\\
            \noalign{\smallskip}                                                
            \hline                                                              
         \end{array}                                                            
     \]                                                                         
   \end{table}                                                                  
                                                                                
\begin{figure}                                                                  
\psfig{figure=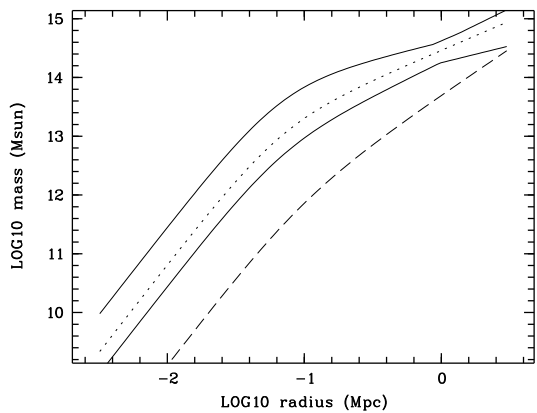,height=6cm}                               
\caption{Mass profile for Cl0024+17. The solid                                  
lines give the range of mass profiles allowed by the combination                
of models while the dotted line gives the best fitting model                    
with an isothermal temperature distribution and a temperature of 6 keV.       
The dashed lower line shows the gas mass profile.                               
}                                                                               
\end{figure}

\section{Discussion and Conclusion}                                                            
                                  
Comparing the gas mass fraction for Cl0024+17  with
the typical values of 20 - 30\% at larger radii for nearby clusters 
(e.g. B\"ohringer 1994, David, Jones, \& Forman 1995, White \& Fabian, 1995) 
we find that the 6 - 8 keV models give quite consistent results.
At temperatures lower than about 5 keV the gas mass fractions are becoming
too high, larger than 35\% (In the sample of White \& Fabian of 19 well
studied nearby clusters for example non of the clusters has an observed 
gas mass fraction larger than 26\% and even the values extrapolated to large
radii never exceed 35\%). 
We should note, however, that the gas masses at the outer radii
are obtained from largely extrapolated X-ray surface brightness profiles.

Compared to the weak lensing mass of $4\cdot 10^{15} h_{50}^{-1}$ M$_{\odot}$
within $3 h_{50}^{-1}$ Mpc (Bonnet et al. 1994) and the virial
mass (Schneider et al. 1986) of 
$6.6\cdot 10^{14} h_{50}^{-1}$ M$_{\odot}$  within $0.48  h_{50}^{-1}$ Mpc  
the mass deduced here from the X-ray observations is lower by a factor of
4. This discrepancy cannot be reconciled by a moderate increase of the gas
temperature, we would rather have to make this the hottest cluster ever
observed to obtain consistency.
The X-ray gas temperature has actually been determined from ASCA observations by
Soucail et al. (1999) with some uncertainty due to contaminating
sources yielding $T_X = 5.5 (+4.5,-1.9)$ keV. Note, that in comparing with
the result on large scales by Bonnet et al., we have extrapolated the 
gas density profile from the observed outer radius of $0.6 h_{50}^{-1}$ Mpc
to $3 h_{50}^{-1}$ Mpc. Since a temperature increase at large radii
is phyically unlikely (see e.g. Markevitch et al. 1998) a larger mass could
be obtained if the slope of the gas density profile steepens significantly.
A steepening of the $\beta $ value by about a factor of 1.5 from the
small value observed at small radii is not impossible. On the other hand
the result by Bonnet et al. (1994) is derived on the assumption of spherical
symmetry up to large radii. Clumping in the mass distribution can help
to reduce the mass required to reproduce the observations. Another
source of uncertainty is the assumed redshift of the lensed objects.
All these effects
could reduce but not remove the discrepancy between the two results.

A comparison with the central lensing masses is more encouraging. 
The lensing mass from the strong lensing model
of Kassiola et al. (1992) and Smail et al. (1997) with $M(R\le 220
h_{50}^{-1}$kpc$) \sim 2\cdot 10^{14} h_{50}^{-1}$ M$_{\odot}$ 
and the weak shear estimate by Smail et al. (1996) with $M(R\le 400
h_{50}^{-1}$kpc$) \sim 2.8 (\pm 0.7) 10^{14} h_{50}^{-1}$ M$_{\odot}$
are roughly consistent with the upper limit of the X-ray results, while 
the result of Tyson et al. (1998) is a bit higher with $M(R\le 220
h_{50}^{-1}$kpc$) \sim 3.2\cdot 10^{14} h_{50}^{-1}$ M$_{\odot}$.
Since in these models the most probable distance to the source was 
generally assumed to be slightly lower than now measured the masses
reduce insignificantly for our discussion by of the order of 10\%.
Broadhurst et al. (1999) who also applied a lens model
for a mass estimate with the newly measured arc redshift find
$M(R \le 200 h_{50}^{-1}$kpc$) \sim 2.22\cdot 10^{14} h_{50}^{-1}$ M$_{\odot}$
very similar to the earlier results by Kassiola et al. (1992) and
Smail et al. (1997).
     
While the X-ray mass may be consistent with the mass of the cluster core, 
there could be much more
mass in an unrelaxed state surrounding the cluster. Thus the cluster could
well be a somewhat scaled-up version of the Virgo cluster for which a core
mass of $1.5 - 6 \cdot 10^{14} $ M$_{\odot}$ has been deduced from X-ray
observations but a much larger mass is indicated by the large diffuse and
irregular X-ray halo (B\"ohringer et al. 1994) and a mass of 
$\sim 10^{15} $ M$_{\odot}$ is deduced from the Virgo infall velocity.

The most interesting morphological result is the small core radius of the
cluster. In some cases equally small core radii in the X-ray 
surface brightness have been measured for other massive clusters 
with cooling flows (e.g. Perseus (Schwarz et al. 1992) or some of the 
the clusters analyzed by Durret et al. 1994 and Mohr et al. 1997).
In this case the central surface brightness peak is
related to the mean temperature drop of the gas in the cooling flow region and
does not necessarily reflect a small core radius of the cluster potential.
This can for example be compared to a large sample of mostly nearby clusters 
analysed by White, Jones, \& Forman (1997) in which the core radius of
the gravitational potential of the clusters was estimated such that
consistent image deprojection and hydrostatic solutions were obtained.
For none of the clusters a core radius smaller than 100 kpc was
implied (the only exception in the sample is the radio galaxy Fornax A
which is not a proper cluster). Smaller  core radii for the distant
lensing clusters have on the contrary often been implied by lensing
studies (e.g. Miralda-Escude 1991, Mellier, Fort, \& Kneib 1993).
In the present case we do not expect a significant influence of 
central cooling. The small core radius is therefore most certainly reflecting
the shape of the gravitational potential. The clusters with small core radii
and cooling flows usually have dominant, central cD galaxies, which is also not found 
for Cl0024+17. 
Therefore it is very assuring that
we recover a very similar core radius as the lensing models of $66 {-25 \brack +38}
h_{50}^{-1}$kpc, while Tyson et al. (1998) find $70 h_{50}^{-1}$ kpc and Smail
et al. (1996) find $40 (\pm 10) h_{50}^{-1}$ kpc. It is probably this small
core radius of the gravitational potential
rather than the overall mass which makes Cl0024+17 such a
spectacular gravitational lens.

\begin{acknowledgements}                                                        
P.S. acknowledges support by the Verbundforschung under grant number 
50 OR 970835. G.S. and Y.M. acknowledge support by the TMR network
``Gravitational Lensing: New Constraints on Cosmology and the Distribution
of Dark Matter'' of the EC under contract No. ERBFMRX-CT97-0172. 

\end{acknowledgements}

\end{document}